\documentclass[prb,aps,preprint,showpacs,superscriptaddress]{revtex4}

\usepackage{graphicx}

\begin{document}

  \title{Electronic and piezoelectric properties of BN nanotubes\\
    from hybrid density functional method}

  \author{H. J. Xiang}
  \affiliation{Hefei National Laboratory for Physical Sciences at
    Microscale, 
    University of Science and Technology of
    China, Hefei, Anhui 230026, People's Republic of China}

  \affiliation{USTC Shanghai Institute for Advanced Studies,
    University of Science and Technology of China,
    Shanghai 201315, People's Republic of China}

  \author{Z. Y. Chen}
  \affiliation{Hefei National Laboratory for Physical Sciences at
    Microscale, 
    University of Science and Technology of
    China, Hefei, Anhui 230026, People's Republic of China}

  \affiliation{USTC Shanghai Institute for Advanced Studies,
    University of Science and Technology of China,
    Shanghai 201315, People's Republic of China}

  \author{Jinlong Yang}
  \thanks{Corresponding author. E-mail: jlyang@ustc.edu.cn}

  \affiliation{Hefei National Laboratory for Physical Sciences at
    Microscale, 
    University of Science and Technology of
    China, Hefei, Anhui 230026, People's Republic of China}

  \affiliation{USTC Shanghai Institute for Advanced Studies,
    University of Science and Technology of China,
    Shanghai 201315, People's Republic of China}

  \date{\today}

  \begin{abstract}
    The electronic and piezoelectric properties of the boron
    nitride (BN) nanotubes are investigated with the hybrid density
    functional (B3LYP) method.
    We first study bulk h-BN and BN
    sheet and find that the B3LYP band structure and energy gap are
    consistent with the GW results. 
    The B3LYP band gap is larger than the LDA one by about 1.8 eV
    for both zigzag and armchair nanotubes with various radius.
    We give an alternative interpretation that the optical
    absorption lines at 4.45 eV might be due to the electron
    transition in small zigzag 
    BN nanotubes. The piezoelectric constant from the B3LYP
    method for zigzag BN nanotubes are substantially larger than
    those in the PVDF polymer family, suggesting BN
    nanotubes as candidates for various nanoelectromechanical
    applications.  
  \end{abstract}

  \maketitle

  Since their discovery in 1991,\cite{cnt1} 
  carbon nanotubes have attracted considerable interest worldwide
  because of their unusual properties and great potentials for
  technological applications. Simple p-band tight-binding model predicts
  that depending on the way of the rolling up the nanotube can be
  metallic or semiconducting or insulating.\cite{cnt2}

  Soon after the discovery of carbon nanotubes it became obvious that
  similar nanostructures could be formed by other elements
  and compounds which form layered structures bearing some
  resemblance to graphite. For example, hexagonal BN (h-BN)
  was predicted on the basis of theoretical 
  calculations\cite{BNNT1} to be capable of forming nanotubes, a
  prediction which was later confirmed experimentally by the synthesis
  of such nanotubes.\cite{BNNT2}
  They are predicted to be semiconductors regardless of diameter,
  chirality, or the number of walls of the tube.\cite{BNNT1} This
  contrasts markedly with the heterogeneity of electronic properties of
  carbon nanotubes, and also makes pure BN nanotubes particularly
  useful for potential device applications. 
  Furthermore, recent experiments indicated that BN nanotubes exhibit a
  stronger resistance to oxidation at high temperatures than carbon
  nanotubes.\cite{BNNT3}
  So as far as the optical and optoelectronic applications of nanotubes
  are concerned, BN nanotubes could be superior to carbon nanotubes.
  An important parameter for optical and optoelectronic applications is
  the optical band gap.
  Although all studies agree that BN nanotubes are
  semiconductors, however, the magnitude of the band gap of BN nanotubes 
  and the band gap dependence on the chirality and radius are on debate
  in the literature. For example, Blase {\it et al.} showed that BN
  nanotubes are wide-gap semiconductors with a constant band
  gap of about 5.5 eV that is independent of the radius and
  helicity by carrying out local-density approximation (LDA)
  and quasiparticles calculations.\cite{BNNT4} 
  A time-dependent localized-density-matrix calculation on BN nanotubes 
  based on a semiempirical Hamiltonian indicated that the optical gap of
  BN nanotubes is independent of the chirality with a given
  tube diameter but dependent on both the tube length and the
  tube diameter.\cite{BNNT5}
  In contrast, two independent LDA calculations show that  though the
  band gap of all the single-walled nanotubes with a diameter larger
  than 15 \AA\ is independent of diameter and chirality, the band gap of
  the zigzag nanotubes with smaller diameters decreases strongly as the tube
  diameters decrease and that of the armchair nanotubes has only a weak
  diameter dependence.\cite{BNNT6,BNNT7} Experimentally, the measured
  band gap for BN nanotubes varies from 4.5 eV to 5.5 eV depending on
  the measurement method and the different synthesized BN
  nanotubes. \cite{BNNT8} Clearly, more theoretical studies employing
  reliable methods are needed to clarify the issue on the band gap of
  BN nanotubes.  

  It is well known that energy gaps between occupied and empty bands provided
  by LDA deviate much more from experimental values.
  One of the successfull methodology to correct the LDA band gap is using
  the GW approximation.\cite{GW1}
  The GW approximation takes account of dynamical screening
  effect of electrons within the random-phase approximation and has been
  applied to a wide range of semiconductors, and turned out
  to improve the band gap significantly to LDA.\cite{GW2}
  The quantum monte carlo (QMC)
  method\cite{QMC1} has also been used to estimate excitation energies
  based on explicitly correlated wavefunctions\cite{QMC2}. 
  However, these calculations are computationally
  very demanding and only applicable to relatively small systems
  in spite of considerable progress made in developing more efficient
  computational algorithms. Recently, the B3LYP hybrid density
  functional method,\cite{B3LYP1,B3LYP2} which is well known in the
  study of thermochemistry of atoms and molecules, has been applied to
  some periodic systems.\cite{B3LYP3} A recent study \cite{B3LYP4} indicated that 
  B3LYP reproduces observed band gaps reliably in a wide variety of
  materials, the B3LYP band gap is at least as accurate as that 
  obtained with sophisticated correlated calculations or
  perturbation theories.

  Recently, electric polarization, piezoelectricity, and
  pyroelectricity in BN nanotubes 
  have attracted much interest.\cite{BN_pz1,BN_pz2,BN_pz3}
  Using LDA calculation, Nakhmanson {\it et al.} showed that
  BN nanotubes are excellent piezoelectric systems with response
  values larger than those of piezoelectric polymers.
  However, previous studies on BeO and ZnO showed that the piezoelectric
  constants depend on the chosen functional: LDA always gives the
  largest absolute value for the piezoelectric constants, whereas the
  lowest absolute value is provided either by HF or B3LYP.
  So the question that how large the piezoelectric constants of BN
  nanotubes can be is opening. Here, in this work, we recalculate
  the piezoelectric constants in BN nanotubes using B3LYP functional.

  In this paper, we study the electronic and piezoelectric properties for
  BN nanotubes using the B3LYP method. In the hybrid functional scheme
  \cite{B3LYP1} the nonlocal 
  Hartree-Fock (HF) approach is mixed into the energy functional of
  the GGA. Here, the Perdew-Wang\cite{pw91} gradient-corrected
  correlation energy, which was used in the original work of
  Becke,\cite{B3LYP1} is replaced by Lee-Yang-Parr correlation
  energy.\cite{B3LYP2} The calculations are carried out with the
  CRYSTAL package.\cite{CRYSTAL} The basis vectors for expanding the
  Kohn-Sham orbitals are linear combinations of atom-centered
  Gaussian basis sets.\cite{basis} The all-electron basis sets
  adapted in the calculations are 6-21G* for B and N.
  The integration in reciprocal space has been carried out using a $1
  \times 1 \times 16$ Monkhorst-Pack k-point mesh to give well
  converged energy. We adopt 7, 7, 7, 7, and 14 as the integral
  tolerances to obtain high precision in monoelectronic and
  bielectronic integrals. The total energy convergence threshold
  exponent is set as 9.

\begin{figure}[!hbp]
    \includegraphics[width=8cm]{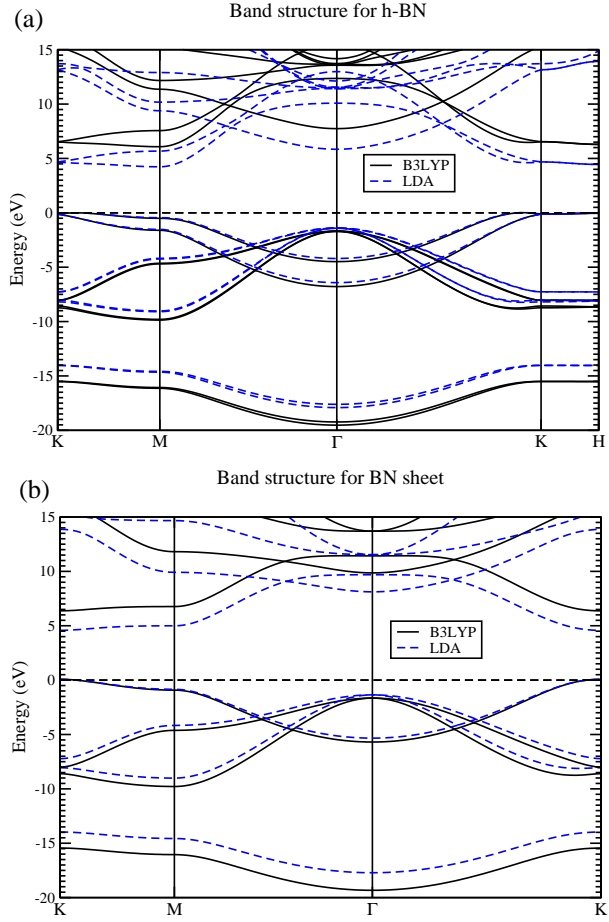}
    \caption{(Color online) B3LYP and LDA band structures of (a) bulk
      h-BN and (b) an isolated BN sheet.}
    \label{fig1}
  \end{figure}

  A single-walled BN nanotube is formed when a piece of hexgonal BN
  (h-BN) sheet is wrapped into a cylindrical form, the edges are
  seamlessly joined together and the ends of the cylinder closed.
  So to get the properties of BN nanotubes, as a first step, we 
  start to study bulk h-BN and an isolated BN sheet.
  The crystalline structure of h-BN is hexagonal and
  has the D$^{4}_{6h}$. It consists of hexagonal
  graphite-like sheets but with an ABAB stacking
  with boron atoms in layer A found directly below
  nitrogen atoms in layer B.
  The experimental lattice parameters \cite{h_BN_exp} ($a=2.504$ \AA\ 
  and $c/a=2.66$) are used to perform all calculations.
  The isolated BN sheet is simulated by a two-dimensional (2D) hexagonal
  slab model with $a=2.504$ \AA.
  Fig.~\ref{fig1} shows the band structures for bulk h-BN and BN sheet.
  In addition, their LDA band structures are also shown in
  Fig.~\ref{fig1}. We find that both B3LYP and LDA predict an indirect
  band gap between the bottom of the conduction band at the $M$ point
  and the top of the valence band near $K$. The indirect and direct
  band gaps for bulk h-BN are shown in Table.~\ref{table1}. The
  agreement between our B3LYP band gaps and previous GW values
  \cite{hBN_GW}is very 
  good. In fact, besides the band gap agreement, we find the B3LYP band
  structure also agrees well with the GW result:
  the difference between B3LYP result and LDA one is strongly dependent 
  on the degree of orbital localization and that the overall
  corrections can not be reproduced by a rigid band shift of the
  conduction states with respect to the top of the valence states.
  For the isolated BN sheet, our results indicate that it is a direct
  gap at $K$ semiconductor with a energy gap of 4.50 eV and 6.30 eV
  within LDA and B3LYP respectively. 
  The good agreement between our B3LYP results and GW ones indicates
  that B3LYP describes very well the band gap and band structure for   
  such systems composed by BN layers. 

\begin{figure}[!hbp]
    \includegraphics[width=8cm]{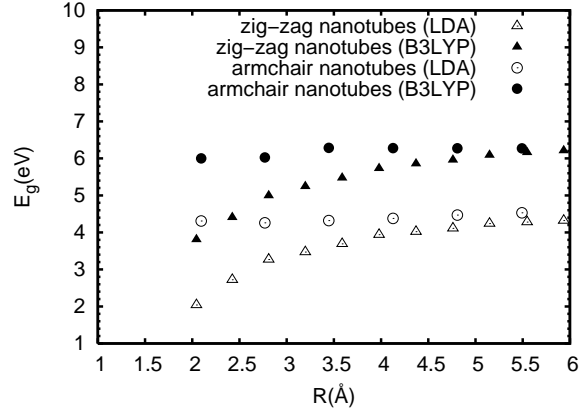}
    \caption{B3LYP band gaps of BN$(n,0)$ zigzag nanotubes with $n=5-15$
      and BN$(n,n)$ armchair nanotubes with $n=3-8$. The radius refers
      to that of an unrelaxed BN nanotube. For comparison, previous LDA
      results are also reproduced here.}
    \label{fig2}
  \end{figure}

\begin{figure}[!hbp]
    \includegraphics[width=8cm]{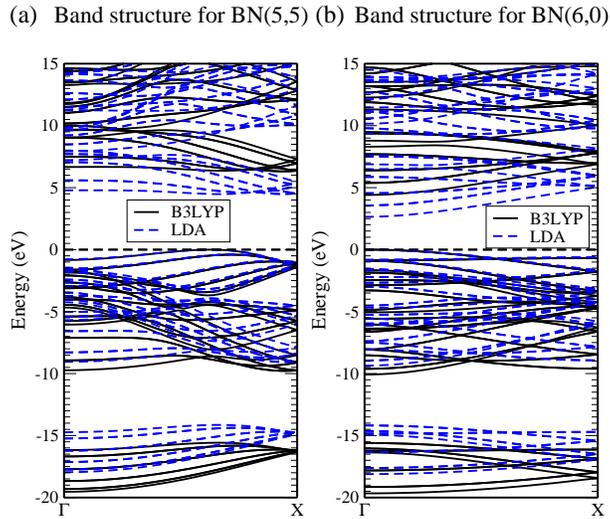}
    \caption{(Color online) B3LYP and LDA band structures of (a)
      armchair BN(5,5) nanotube and (b) zig-zag BN(6,0) nanotube.}
    \label{fig3}
  \end{figure}

  Now we turn to study BN nanotubes. 
  First we fully optimized the zigzag BN$(n,0)$ nanotubes with $n=5-15$
  and armchair BN$(n,n)$ nanotubes with $n=3-8$ using the B3LYP
  functional. The relaxed geometries for BN nanotubes are similar with
  previous LDA calculations.\cite{BNNT6} LDA is well known to predict
  good structural parameter but strongly underestimate the band gap for
  semiconductors. Here we discuss mainly the electronic
  properties obtained from the B3LYP calculations. We plot the band gap
  dependence upon radius of BN nanotubes in Fig.~\ref{fig2}. First, the
  B3LYP band gap dependence is almost the same as previous LDA results:
  The energy gap of small zigzag BN nanotubes decreases rapidly with
  the decrease of radius, in constrast, the energy gap armchair BN
  nanotubes has weak radius dependence. Secondly, the difference in the
  band gap between B3LYP band gap and LDA one is almost a constant
  (about 1.8 eV) for both zigzag and armchair nanotubes with various
  radius. Thirdly, For BN nanotubes with radius larger than 6 \AA\, the
  B3LYP band gap is about 6.20 eV, agreed well with the B3LYP band gap
  for the isolated BN sheet. In a very recent study on BN nanotubes,
  three optical absorption lines were observed by means of optical
  absorption spectroscopy.\cite{BN_opt} The absorption line at 5.5 eV was
  attributed to the transitions between pairs of van Hove
  singularities in the one-dimensional density of states of BN 
  single-wall nanotubes. The low energy line at 4.45 eV was considered
  to caused by the existence of a Frenkel exciton with a binding
  energy in the 1 eV range. From our calculation, the absorption line
  at 4.45 eV might also result from the transitions between pairs
  of van Hove singularities of BN nanotubes, such as BN(6,0) nanotube.  
  Since B3LYP gives almost the same band
  structure for bulk h-BN as the GW quasiparticle one, it is expected that
  B3LYP would also give the accurate band structures for BN nanotubes. 
  Thus we present the band structures for BN(5,5) and BN(6,0) nanotubes in
  Fig.~\ref{fig3}. Both B3LYP and LDA results are shown.
  We can see that the difference between B3LYP band LDA band structures
  can't be described by just raising the conduction band by the
  band gap difference. Not only the conduction bands are shifted
  upwards, but also the B3LYP valence bands below 3 eV with respect to
  the valence top differ significantly from the LDA counterpart: The
  most obvious difference is the B3LYP bands at about $-15$ eV below the
  valence top shift downwards 1.5 eV.   

\begin{figure}[!hbp]
    \includegraphics[width=8cm]{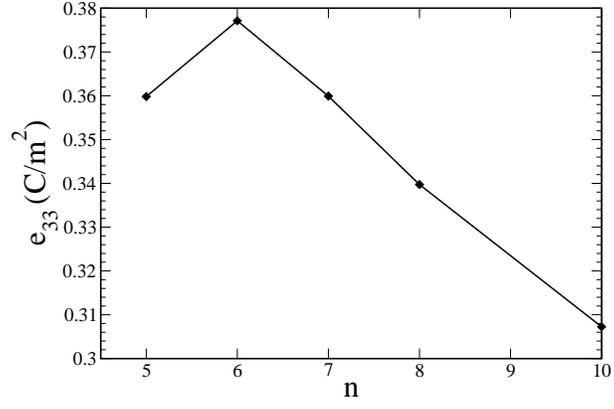}
    \caption{Piezoelectric constant
      for zig-zag BN($n,0$) nanotubes assembled in a bundle assuming a
      close packed geometry with intertube equilibrium distance of 3.2 \AA.}
    \label{fig4}
  \end{figure}

  In the last decade, it has become possible to evaluate the
  components of the piezoelectric tensor and other dielectric
  properties such as the spontaneous polarization and the effective
  Born charges through a technique based on the evaluation
  of one Berry phase.\cite{berry} 
  The Berry-phase method can be employed to compute
  piezoelectric properties, which are directly 
  related to polarization differences between strained and unstrained 
  tubes. As most BN nanotubes synthesized experimentally are of
  zig-zag type,\cite{BN_zigzag} we only study the piezoelectric properties of
  zig-zag BN nanotubes. In the calculations, we model the isolated BN
  nanotubes with the supercell large enough to prevent the neighbour BN
  walls from interacting.
  However, since piezoelectric constants are well defined
  for three-dimensional systems and in order to better compare with previous
  LDA results, the calculated piezoelectric constants are normalized
  using the ratio between the volume of the supercell and that of the
  bundle of nanotubes assuming a close packed geometry with
  intertube equilibrium distance of 3.2 \AA.
  For zig-zag BN nanotubes, 
  the only surviving piezoelectric strain tensor component is 
  $e_{33}=\delta P_{3}/\epsilon_{3}$, where $\epsilon_{3}=c-c_{0}/c_{0}$ 
  ($c$ is the lattice constant and $c_{0}$ is equilibrium values of $c$).
  The piezoelectric constant for different zigzag BN nanotubes are
  plotted in Fig.~\ref{fig4}. If we assume that one BN pair contributes
  the same to $e_{33}$ for all zigzag BN nanotubes with various radius, 
  $e_{33}$ should be approximately inversely proportional to the radius of
  the BN nanotube. It is indeed the case for $n>5$, however, there is an
  abnormal small decrease of $e_{33}$ for BN(5,0) nanotube. We attribute
  the abnormal phenomenon to the strong $\sigma-\pi$ hybridization in
  BN(5,0) nanotube.  
  Comparing with the LDA results obtained by Nakhmanson {\it et al.}, 
  the current B3LYP piezoelectric constants have similar magnitude.
  However, the LDA piezoelectric constants increase along with the
  decrease of the radius, in other words, BN(5,0) nanotube has the
  largest piezoelectric constant for all BN nanotubes studied.
  The discrepancy between our B3LYP and previous LDA results may result
  from the different functionals employed.
  As mentioned above, previous studies implied that piezoelectric
  constants for ZnO and BeO 
  calculated using different Hamiltonians differ substantially. 
  Here, we find that within both LDA and B3LYP, the piezoelectric
  constants for zig-zag BN nanotubes have similar magnitude,
  substantially larger than those (about $0.12$ C/m$^{2}$) in the PVDF
  polymer family. 

  By employing the B3LYP functional, we examine the electronic
  and piezoelectric properties of BN nanotubes. 
  The applicability of the B3LYP method in such BN systems is illustrated
  by the similarity between the B3LYP band structure of bulk h-BN and
  previous GW quasi-particle band structure. 
  The current work supports previous LDA results that the energy gap
  of small zigzag BN nanotubes decreases rapidly with the decrease of
  radius, whereas, the energy gap of armchair BN nanotubes almost
  remains constant.  
  The obtained B3LYP band gap is larger than the LDA one by almost constant
  (about 1.8 eV) for nanotubes with various chirality and radius.
  We give an alternative interpretation that the absorption lines
  at 4.45 eV might be due to the electron transition in small zigzag
  BN nanotubes. 
  The piezoelectric constant within the B3LYP
  formalism for zigzag BN nanotubes are substantially larger than
  those in the PVDF polymer family, thus piezoelectric BN nanotubes
  hold promise for application in nanometer scale sensors and
  actuators. 

  This work is partially supported by the National Project for the
  Development of Key Fundamental Sciences in China (G1999075305), by
  the National Natural Science Foundation of China 
  (50121202, 10474087), by the USTC-HP HPC project, by 
  the EDF of USTC-SIAS, and by the SCCAS.

  \newpage

  \begin{table}
    \caption{Band gaps of bulk h-BN} 
    \begin{tabular}{ccccc}
      \hline
      \hline
      E$_{g}$ (eV)  & B3LYP & Our LDA & Others' LDA & GW \\
      \hline
      direct & 6.33    & 4.48 & 4.46   & 6.47       \\
      indirect & 6.06  & 4.22 & 4.02   & 5.95      \\
      \hline
      \hline
    \end{tabular}
    \label{table1}
  \end{table}

\end{document}